\documentclass[prl,twocolumn,showpacs,aps,superscriptaddress,preprintnumbers,reprint]{revtex4-2}
\usepackage [dvips]{graphicx}
\usepackage{amsmath}\usepackage{amsfonts}
\usepackage{mathrsfs}
\usepackage{color}
\definecolor{darkblue}{rgb}{0,0,0.6}
\usepackage[colorlinks,linkcolor=darkblue,citecolor=darkblue,urlcolor=darkblue]{hyperref}

\newcommand{\beeta}{\boldsymbol{\eta}}

\newcommand{\bA}{\text{\bf A}}

\newcommand{\dd}{\text{d}}

\newcommand{\ee}{\text{e}}

\newcommand{\p}{\partial}

\newcommand{\br}{\text{\bf r}}
\newcommand{\be}{\text{\bf e}}
\newcommand{\eps}{\varepsilon}
\newcommand{\bk}{\text{\bf k}}

\newcommand{\bq}{\text{\bf q}}

\newcommand{\bnabla}{\boldsymbol{\nabla}}

\usepackage[percent]{overpic}
\definecolor{v}{rgb}{0.33,0.67,0}
\definecolor{redUP}{RGB}{138,21,56}
\definecolor{blueUP}{RGB}{21,56,138}
\definecolor{greenUP}{RGB}{56,138,21}
\definecolor{gris}{RGB}{160,160,160}

\begin{document}

\title{Sampling efficiency of transverse forces in dense liquids}

\author{Federico Ghimenti}

\affiliation{Laboratoire Mati\`ere et Syst\`emes Complexes (MSC), Université Paris Cité  \& CNRS (UMR 7057), 75013 Paris, France}

\author{Ludovic Berthier}

\affiliation{Laboratoire Charles Coulomb (L2C), Université de Montpellier \& CNRS (UMR 5221), 34095 Montpellier, France}

\affiliation{Yusuf Hamied Department of Chemistry, University of Cambridge, Lensfield Road, Cambridge CB2 1EW, United Kingdom}

\author{Grzegorz Szamel}

\affiliation{Department of Chemistry, Colorado State University, Fort Collins, Colorado 80523, United States of America}

\author{Fr\'ed\'eric van Wijland}

\affiliation{Laboratoire Mati\`ere et Syst\`emes Complexes (MSC), Université Paris Cité  \& CNRS (UMR 7057), 75013 Paris, France}

\date{\today}

\begin{abstract}
Sampling the Boltzmann distribution using forces that violate detailed balance can be faster than with the equilibrium evolution, but the acceleration depends on the nature of the nonequilibrium drive and the physical situation. Here, we study the efficiency of forces transverse to energy gradients in dense liquids through a combination of techniques: Brownian dynamics simulations, exact infinite-dimensional calculation and a mode-coupling approximation. We find that the sampling speedup varies non-monotonically with temperature, and decreases as the system becomes more glassy. We characterize the interplay between the distance to equilibrium and the efficiency of transverse forces by means of odd transport coefficients.
\end{abstract}

\maketitle

To sample a given target distribution, the paradigm is to construct Markov processes endowed with detailed balance. When the physics slows the dynamics down, as for instance in the vicinity of a critical point, or in disordered and dense systems, algorithms that can increase the sampling efficiency are much needed~\cite{krauth2021event, berthier2023modern}. Sampling by violating detailed balance using nonequilibrium dynamics is a possible route, explored in an applied mathematics literature dating back to the mid-nineties~\cite{hwang1993accelerating, chen1999lifting, diaconis2000analysis, vucelja2016lifting}. Potential applications are not limited to physical systems, since, for instance, slow dynamics caused by a complex non-convex energy landscape are also encountered in machine learning and neural networks~\cite{Ohzeki_2016, gao2020breaking, futami2020accelerating, futami2021accelerated}. Bounds and inequalities on the convergence or mixing rates have been obtained~\cite{franke2010behavior, ichiki2013violation, lelievre2013optimal, rey2015irreversible, hwang2015variance, duncan2016variance, levin2017markov,apers2017does}, and studies encompass the mean-field Ising model~\cite{monthus2021large} and systems evolving via diffusive hydrodynamics~\cite{kaiser2017acceleration, kaiser2018canonical, coghi2021role}. This is a very active field of applied mathematics~\cite{chatterjee2021correction, franke2021note} and of computer science~\cite{apers2017lifting, apers2017fast,  apers2018simulation, apers2021characterizing}. Numerical studies also exist for a variety of systems~\cite{bernard2009event, bernard2011two,kapfer2017irreversible,isobe2015hard,ohzeki2015langevin,michel2015event,sakai2013dynamics,turitsyn2011irreversible,ghimenti2022accelerating}, but no quantitative results exist for systems with self-induced disorder, such as glassy liquids. In the latter case, nonequilibrium forces can either shift~\cite{berthier2013non} or destroy~\cite{berthier2000two} the glass transition, while the addition of unphysical degrees of freedom was recently shown~\cite{ninarello2017models} to drastically change the relaxation dynamics. 

We explore how nonequilibrium methods that sample the Boltzmann distribution fare when applied to a strongly-interacting classical many-body system, such as a high-density or low-temperature fluid exhibiting glassy dynamics, and determine the dependence of the acceleration on the state point. The specific dynamics we study is the overdamped Langevin dynamics driven out of equilibrium by a force field transverse to the local energy gradient. Our results are established using a combination of techniques, ranging from the numerical integration of Langevin equations for a Kob-Andersen mixture, through mean-field infinite dimensional calculation to finite-dimension mode-coupling approximation. Our presentation goes along these three axes, each of which sheds its own light on the questions we ask.

\begin{figure}[b]
    \includegraphics[width = \columnwidth]{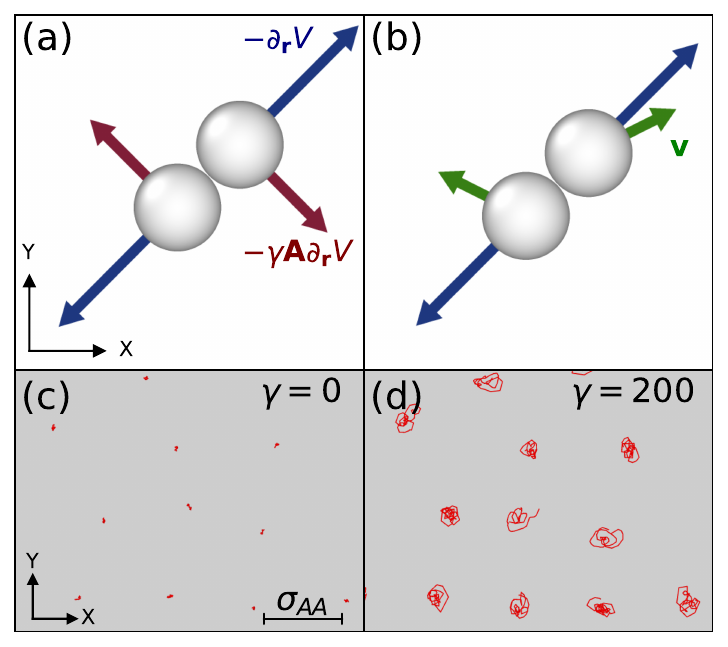}
    \caption{(a) The addition of transverse forces (red) to the  potential ones (blue) acts similarly to (b) a dynamics lifted by additional degrees of freedom (green), by creating an effective chirality. (c) Rendering of a short trajectory for a few particles without any transverse force at $T=0.8$. (d) Same with transverse forces which induce circular trajectories.}\label{fig:cartoon0}
\end{figure}

We demonstrate the existence of an optimal temperature for the acceleration. While a gain remains, the efficiency decreases as the glass transition is approached. Transverse forces also lead to the appearance of odd transport coefficients, that were earlier found in active matter systems composed of chiral particles or driven by nonreciprocal forces~\cite{hargus2021odd,poggioli2023odd,banerjee2017odd} (see also \cite{kalz2022collisions,kalz2023oscillatory} for a dilute equilibrium fluid). Surprisingly, odd diffusivity is insensitive to the emergence of glassy behavior.

Our approach is illustrated with the example of a single particle with position $\br$ in an external potential $V(\br)$ at temperature $T=\beta^{-1}$,
\begin{equation}\label{eq:deftf}
\frac{\dd\br}{\dd t}=-\mu(1+\gamma {\mathbf A})\p_\br V+\sqrt{2\mu T}\beeta ,
\end{equation}
where $\mu$ is the mobility. The components of the Gaussian white noise $\beeta$ are independent, $\langle \eta_i(t)\eta_j(t')\rangle=\delta_{ij}\delta(t-t')$. When the matrix ${\mathbf A}$ is skew-symmetric, the nonequilibrium force of strength $\gamma$ is transverse to the energy gradient. The stationary distribution thus retains its Boltzmann form, $\rho_\text{B}(\br)=\ee^{-\beta V(\br)} / Z$ even when $\gamma>0$, but the entropy production rate is finite, $\tau_\Sigma^{-1}= \beta \gamma^2\langle ({\mathbf A}\p_\br V)^2\rangle_\text{B}$. Another relevant time scale governing microscopic dynamics is given by the reciprocal of the average escape rate~\cite{pitard2011dynamic,fullerton2013dynamical,MAES20201}. Its equilibrium expression is $\tau_0^{-1}\sim\langle\beta(\partial_\br V)^2\rangle_\text{B}$ but with a nonzero $\gamma$ this becomes $\frac{\tau_0}{1+\gamma^2||A||^2_\text{F}/d}$ (see SM~\cite{SM}). This elementary reasoning would suggest that transverse forces simply result in a global rescaling of time scales. We show below that many-body interactions lead to a very different picture. 

To widen the scope of our statements, we highlight a correspondence between transverse forces, as defined in Eq.~\eqref{eq:deftf}, and the lifting procedure~\cite{chen1999lifting, diaconis2000analysis, vucelja2016lifting,chatterjee2021correction,franke2021note}, which is an alternative approach to accelerate the dynamics. In a nutshell, lifting amounts to augmenting the degrees of freedom of a system with equilibrium dynamics by a set of auxiliary (and unphysical) variables that produce nonequilibrium flows while preserving the original equilibrium distribution. One way to see the connection with transverse forces is to consider, following \cite{ohzeki2015langevin}, two equilibrium systems with potentials  $V_1(\br_1)$ and $V_2(\br_2)$ evolving through the coupled dynamics
\begin{equation}\begin{split}
    \frac{\dd\br_1}{\dd t}=\mu \left(-\p_{\br_1}V_1+\gamma\p_{\br_2} V_2\right)+\sqrt{2 T \mu}\beeta_1,\\
    \frac{\dd\br_2}{\dd t}=\mu\left(-\p_{\br_2}V_2-\gamma\p_{\br_1} V_1\right)+\sqrt{2 T \mu}\beeta_2.
\end{split}\end{equation}
In this case, the nonequilibrium forces are transverse in the extended $(\br_1,\br_2)$ space and the stationary distribution decouples into $\rho_B(\br_1,\br_2)\propto \ee^{-\beta V_1(\br_1)}\ee^{-\beta V_2(\br_2)}$. If we choose for system 2 a quadratic potential, the equation of motion for 1 resembles that of an active Ornstein-Uhlenbeck particle~\cite{koumakis2014directed} where the role of the self-propulsion velocity is played by $\br_2$~\cite{SM}. The equation of motion for 2 is however different from the Ornstein-Uhlenbeck equation to ensure that 1 samples the Boltzmann distribution: the dynamics of system 1 is lifted by that of system 2. A cartoon of the connection between transverse forces and lifting is shown in Fig.~\ref{fig:cartoon0}. While our derivations start from transverse forces, our conclusions may therefore extend to some locally lifted systems. 

\begin{figure}[t]
\includegraphics[width=\columnwidth]{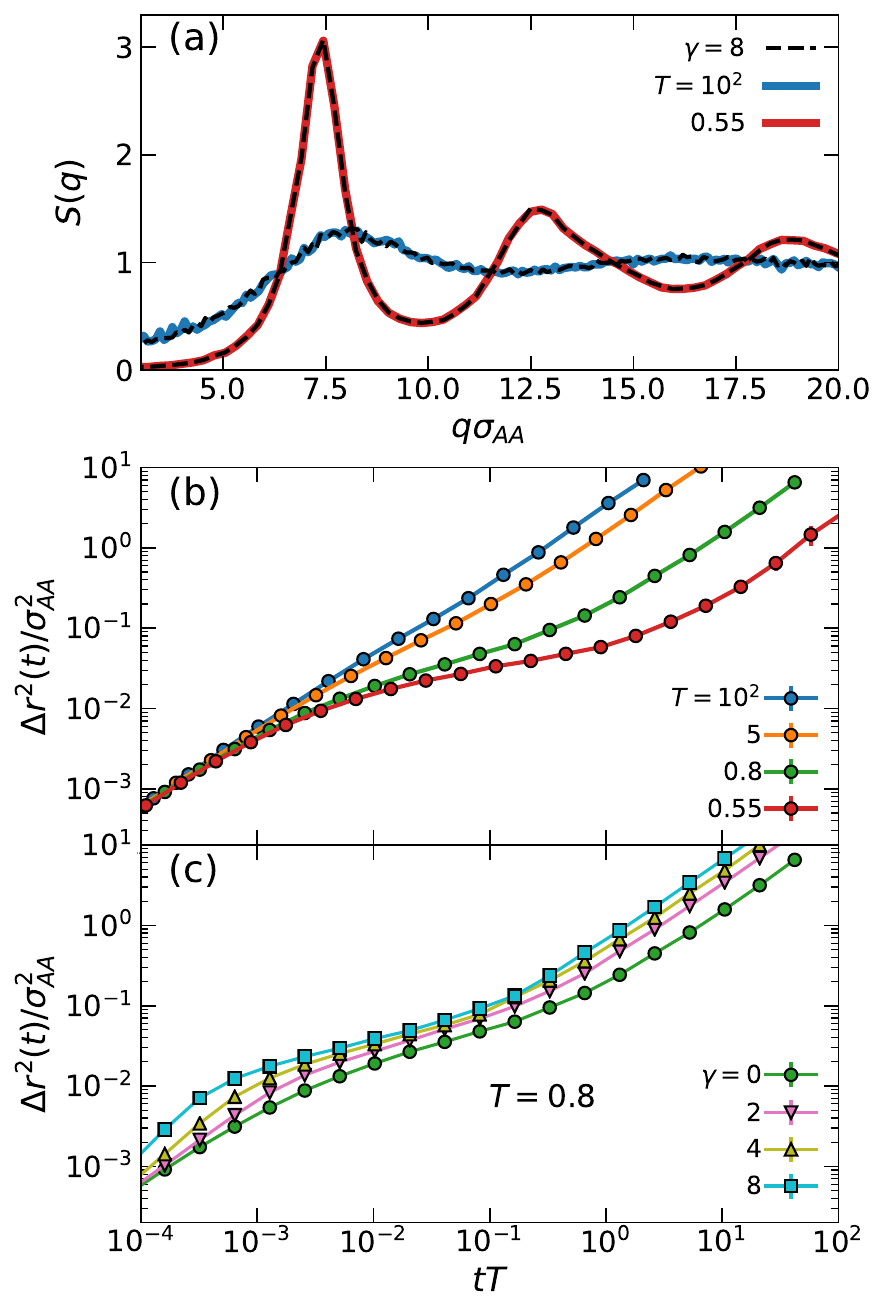}
\caption{(a): Static structure factor at $\gamma=0$ (full color) and $\gamma=8$ (dashed) at two temperatures. (b): Mean-squared displacement of the $A$ particles for different temperatures. Two-step dynamics becomes visible below the onset temperature near $T \approx =1.0$. (c) Mean-squared displacement of the $A$ particles at $T=0.8$ for various values of $\gamma$.}
\label{fig:diffusion}
\end{figure}

The general many-body problem considered is a system with $i=1,\ldots,N$ particles in $d$ dimensions evolving under the influence of interparticle forces ${\bf F}_i=-({\bf 1}+\gamma {\bf A}) \sum_{j\neq i}\p_{\br_i} V(\br_i-\br_j)$, where $\mathbf A$ is a skew-symmetric matrix, and $V({\bf r})$ is a pair potential, evolving as 
in Eq.~\eqref{eq:deftf} with thermal noise. The strength of the nonequilibrium forces is controlled by $\gamma$ which means we keep the matrix elements of $\mathbf A$ or order $1$ (and independent of $\gamma$). The steady state distribution is again the Boltzmann distribution $\rho_{\text{B}} \propto e^{-\beta \sum_{i<j}V(\br_i - \br_j) }$. We first report the results of numerical simulations of a three-dimensional binary Kob-Andersen mixture~\cite{kob1994scaling,kob1995testingI,flenner2005relaxation} of $N_A=800$ particles of type $A$ and $N_B=200$ of type $B$ interacting as
\begin{equation}
V_{\alpha\beta}(r)=4\eps_{\alpha\beta}\left[\left(\frac{\sigma_{\alpha\beta}}{r}\right)^{12}-\left(\frac{\sigma_{\alpha\beta}}{r}\right)^{6}\right],\,r\leq 2.5\sigma_{\alpha\beta}
\end{equation}
with $\alpha,\beta\in\{A,B\}$ and where $\eps_{AA}=1$, $\eps_{AB}=1.5$, $\eps_{BB}=0.5$, $\sigma_{AA}=1$, $\sigma_{AB}=0.8$, $\sigma_{BB}=0.88$. The linear size of the system is $9.4\sigma_{AA}$ and periodic boundary conditions were used. We choose $\mathbf A$ in a block diagonal form  with $\pm 1$ elements in the $xy$ plane, without loss of generality. We integrate the equations of motion using a Euler-Heun algorithm with a discretization step calibrated to optimize efficiency while still properly sampling equilibrium properties~\cite{gleim1998how}. We first show in Fig.~\ref{fig:diffusion}(a) that the static structure is unaffected by the introduction of transverse forces, demonstrating equilibrium sampling with nonequilibrium dynamics, over a temperature range encompassing a high-temperature almost structureless fluid down to a mildly supercoooled liquid. To estimate the speedup of the sampling, we use the mean-squared displacement $\Delta r^2(t)$ for particles A. Its temperature evolution is shown in Fig.~\ref{fig:diffusion}(b) at equilibrium, which displays the development of a two-step glassy dynamics below the onset temperature near $T \approx 1.0$. 

\begin{figure}[t]
\includegraphics[width=\columnwidth]{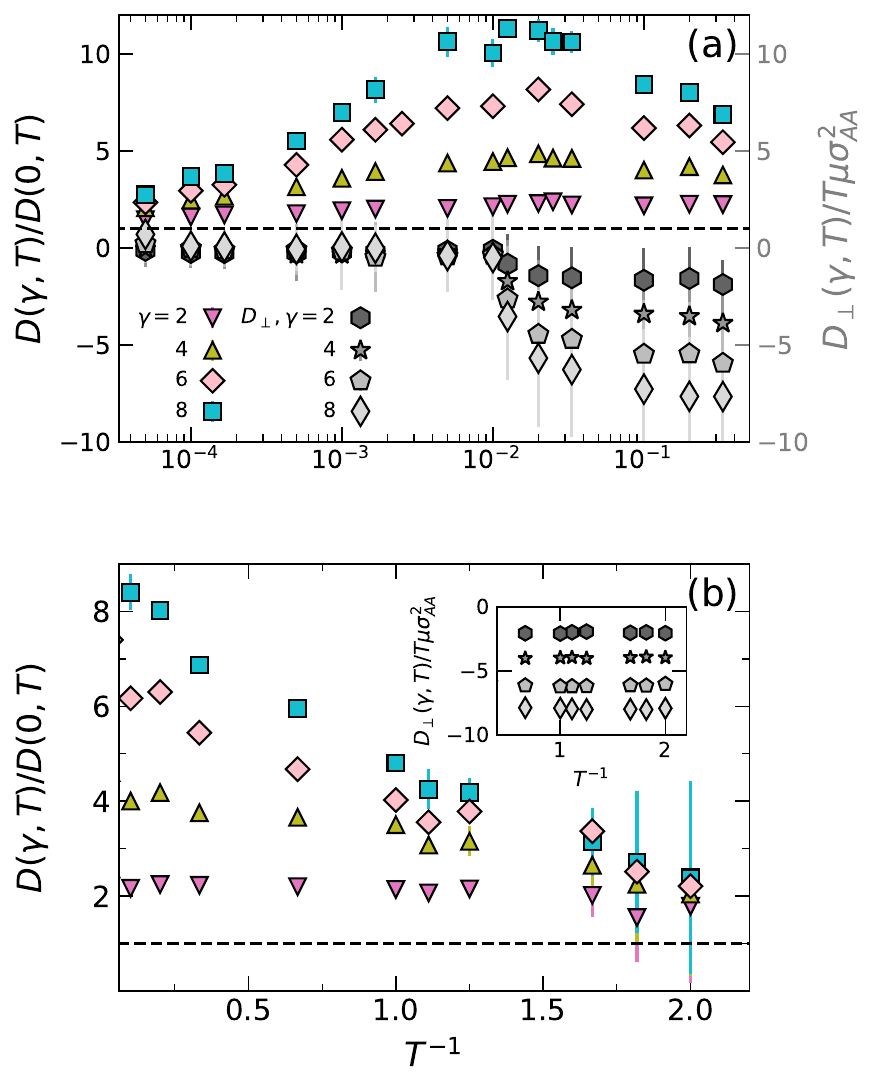}
\caption{(a) Diffusion constant $D(\gamma,T)$ normalised by its equilibrium value at $\gamma=0$ as a function of inverse temperature. The temperature axis uses a logscale to emphasize the non-monotonic dependence. 
The right axis describes the odd-diffusivity as a function of $T^{-1}$ for various values of $\gamma$.
(b) Same as (a) using a linear scale to concentrate on the glassy regime below $T=1.0$. The black dashed line corresponds to the equilibrium efficiency.}\label{fig:D_gamma_T}
\end{figure}

In Fig.~\ref{fig:diffusion}(c), we demonstrate that the introduction of transverse forces accelerates the dynamics of the system. To quantify this acceleration, we extract the diffusion constant, $D(\gamma,T)$, from the long-time limit of the mean-squared displacements, see Fig.~\ref{fig:D_gamma_T}(a). At fixed $\gamma$, there exists a temperature near $T^* \approx 100$ that maximizes the increase of the diffusion constant. 

At high temperatures, interactions (including chiral ones) are smeared out by thermal noise which degrades the efficiency. The initial increase of the acceleration is then well captured by a weak fluctuation expansion~\cite{federico-mct}. The drop of acceleration as the temperature is lowered can be rationalized by the fact that the energy landscape remains unaffected by the transverse forces. When the supercooled regime is entered more deeply,  particles spin along circular trajectories within their local cages, see Fig.~\ref{fig:cartoon0}(c,d). This local motion has a modest influence on the long-time dynamics. We thus expect that the glass transition occurs at the same temperature as in equilibrium.  These two opposite trends account for the existence of an acceleration maximum. 

To confirm the picture of a swirling motion inside a local cage we measure the odd diffusivity of the particles A, which can be calculated from a Green-Kubo expression~\cite{hargus2021odd}
\begin{equation}
D_\perp=\frac{1}{2N_A} \sum_{i=1}^{N_A} \int_0^{+\infty}\dd t\left\langle\dot{y_i}(t)\dot{x_i}(0)-\dot{x_i}(t)\dot{y_i}(0)\right\rangle .
\end{equation}
By symmetry, $D_\perp$ vanishes for an equilibrium dynamics, and its value usefully quantifies the circular motion shown in Fig.~\ref{fig:cartoon0}(d). For example, in the limiting case of a particle trapped in a harmonic well, we show in SM~\cite{SM} that $D_\perp=-\mu\gamma T$, with $\mu$ the mobility of the particle.  

The temperature dependence of $D_\perp$ for our simulated system is shown in Fig.~\ref{fig:D_gamma_T}(a) for different values of $\gamma$. Its absolute value increases with $\gamma$. At fixed $\gamma$, the odd diffusion starts from $0$ at high $T$: as thermal fluctuations wash out interactions, they also suppress particles' chiral motion, which is induced by transverse forces. The modulus of $D_\perp$ rises then steeply as a function of $T^{-1}$ from $0$ to a finite value near $T^*$. As the system enters its slow dynamical regime, $D_\perp$ settles to a finite value as shown in the inset of Fig.~\ref{fig:D_gamma_T}(b). Interestingly, the observed behavior in the arrested glass phase, where it is presumably dominated by the in-cage circular motion created by the transverse forces, agrees with the predictions for the harmonic well. This contrasts with the translational diffusion coefficient which changes by orders of magnitude in the supercooled liquid, and vanishes in the glass. 

Overall the simulations reveal a non-monotonic temperature dependence of the sampling efficiency of transverse forces, which decreases when temperature is lowered, accompanied by odd-diffusivity which appears insensitive to this evolution. To understand these non-trivial findings, we turn to two analytical approaches, focusing for simplicity on a monodisperse fluid. 

First, we consider the mean-field limit which is achieved, for simple fluids, by increasing the space dimension $d$ to infinity~\cite{frisch1985classical,parisi2010mean,maimbourg2016solution,parisi2020theory,liu2021dynamics}, while keeping the number of neighbors per space direction of order unity. In this limit, one can derive an effective Langevin equation for  the position of a tagged particle with an effective noise that originates from the remaining components and the coordinates of all other particles. We defer technical details to \cite{federico-infinite}. Even out of equilibrium~\cite{agoritsas2019out1,agoritsas2019out2}, the influence of the bath appears as a sum of a position-independent friction kernel and a noise. The friction kernel and the noise autocorrelation need to be determined self-consistently, in a typical mean-field procedure. For our problem, we find that the position $\br_i$ of tagged particle $i$ evolves according to
\begin{eqnarray}
 \frac{\dd \br_i(t)}{\dd t}=-\mu\beta\int\dd t' (\mathbf{1}+\gamma {\bf A}){\mathbf M}(t-t')\frac{\dd\br_i(t')}{\dd t'}+\boldsymbol{\Xi}_i(t),
\end{eqnarray}
where ${\mathbf M}$ is a $d\times d$ memory kernel to be determined, and $\boldsymbol{\Xi}_i$ is a zero-average Gaussian noise with correlations
\begin{equation}\begin{split}
\langle \boldsymbol{\Xi}_i(t)\otimes\boldsymbol{\Xi}_j(t')\rangle=&\delta_{ij}\Big[2T\mu \mathbf{1}\delta(t-t')\\
&+(\mathbf{1}+\gamma {\bf A}){\mathbf{M}}(t-t')(\mathbf{1}-\gamma {\bf A})\Big].
\end{split}
\end{equation}
The memory kernel ${\mathbf M}$ is given by correlations of the pair-potential gradients,
\begin{equation}\label{eq:sceq}
{\mathbf M}(t)=\sum_j \langle \bnabla_i V(\br_{ij}(t))\otimes \bnabla_i V(\br_{ij}(0)\rangle,
\end{equation}
where $\br_{ij}=\br_i-\br_j$. To determine ${\mathbf M}$ we need to consider the evolution of the relative position $\br=\br_{ij}$, which can be shown~\cite{federico-infinite} to follow 
\begin{equation}\label{eq:relative}
\frac{1}{2} \frac{\dd \br}{\dd t}=-\mu(\mathbf{1}+\gamma\mathbf{A})\p_\br V-\frac{\mu\beta}{2}\int\dd t' (\mathbf{1}+\gamma {\bf A}){\mathbf M}(t-t')\frac{\dd\br}{\dd t'}+\boldsymbol{\Xi}(t)
\end{equation}
where the Gaussian noise $\boldsymbol{\Xi}$ has correlations $\langle\boldsymbol{\Xi}(t)\otimes\boldsymbol{\Xi}(t')\rangle=T\mu \delta(t-t')+\frac 12 (\mathbf{1}+\gamma {\bf A}){\mathbf{M}}(t-t')(\mathbf{1}-\gamma {\bf A})$. The procedure is to determine the statistics of $\br$ as a functional of ${\bf M}$, and then to determine the force statistics in the rhs of Eq.~\eqref{eq:sceq} as a functional of $\bf M$, hence obtaining a self-consistent functional equation for $\bf M$. In practice, even for equilibrium dynamics, $\bf M$ can only be determined numerically~\cite{manacorda2020numerical}.  To evaluate the diffusion constant we only need the time integral of the kernel which becomes diagonal, ${\bf M}={\mathbf 1}M$. The diffusion constant is expresssed in terms of its time integral $\widehat{M}(\gamma, T)=\int_0^{+\infty}M(t)\dd t$ as
\begin{equation}
\label{eq:Ddinfinie}
D(\gamma,T)=T\mu\frac{1+(1+\gamma^2)\beta\widehat{M}}{(1+\beta\widehat{M})^2+(\beta\gamma\widehat{M})^2}.
\end{equation}
This result is obtained in even space dimension for a matrix $\bf A$~\cite{federico-infinite} made of $d/2$ identical $2\times 2$ blocks with $\pm 1$ entries (nonidentical blocks would require averaging over the blocks, without affecting our conclusions; working in an odd space dimension would involve a single extra space dimension with negligible effect as $d\gg 1$). In the ergodic phase, one can show that the $\gamma$-dependent relaxation time of ${M}$ scales as $\tau(\gamma)=\widehat{M}\propto \gamma^{-1}$ when $\gamma\gg 1$. The diffusion constant $D(\gamma,T)$ therefore behaves as $D \sim \gamma$ for large $\gamma$. Note also that when $M(t)$ does not relax to $0$ then $\widehat{M}=+\infty$ and $D(\gamma,T)$ vanishes. We expect that the constraint of Boltzmann sampling is so strong that transverse forces cannot prevent the emergence of diverging free energy barriers leading to ergodicity breaking, in contrast with active forces~\cite{berthier2013non} or shear flows~\cite{berthier2000two}.  This implies that the dynamical transition temperature at nonzero $\gamma$ is unaffected by the transverse forces~\cite{federico-infinite}. 

The quantity $\widehat{M}$ diverges at a finite temperature while a low density approximation shows~\cite{federico-infinite} that it  increases with $1/T$. It is thus natural to expect that $\widehat{M}$ is an increasing function of $1/T$. Under this assumption, and looking at Eq.~\eqref{eq:Ddinfinie}, we see that $D$ becomes a nonmonotonous function of temperature. Since the maximum of $D$ occurs at high temperature where memory is weak, it makes sense to evaluate Eq.~\eqref{eq:Ddinfinie} using a low density approximation for $\widehat{M}$. It turns out that equilibrium expression of $\widehat{M}$ obtained in \cite{manacorda2020numerical} still holds at nonzero $\gamma$~\cite{federico-infinite}, and it produces an evolution of $D$ consistent with Fig.~\ref{fig:D_gamma_T}(a) as explicitly shown in SM~\cite{SM}.

We also obtain the odd diffusivity, given by
\begin{equation}
D_\perp(\gamma,T)=-\gamma T\mu \frac{\beta \widehat{M}+(1+\gamma^2)(\beta\widehat{M})^2}{(1+ \beta \widehat{M})^2+(\gamma \beta \widehat{M})^2} ,
\end{equation}
and which behaves as $D_\perp \sim \gamma$ at large $\gamma$. At very high temperature we have $D_\perp=-\gamma\mu \widehat{M}\simeq 0$. We also see that $D_\perp=-\gamma\mu T$ below the dynamical transition temperature, when $\widehat{M} \to \infty$, consistently with the harmonic well picture~\cite{SM}. In the mean-field limit, a genuine glass phase appears at low temperature in which particles are trapped in a local harmonic environment created by their neighbors. In a harmonic well  the spectrum of the Fokker-Planck operator only picks up an imaginary part when transverse forces are applied, leaving the real part of the eigenvalues unchanged~\cite{SM}, thereby capturing the emergence of circular orbits within the well. The physical picture is that chiral forces eventually lose their accelerating power by wasting the injected energy into circular trajectories.

It is unclear whether these mean-field results are valid in finite dimensions. We thus resort to an approximate theory in the spirit of the mode-coupling theory of glassy dynamics~\cite{gotze2009complex}. To compare with the infinite-dimensional calculation we focus on the self-part of the intermediate scattering function, $F_s(\bq,t)=\frac{1}{N}\sum_{j} \langle e^{i\bq\cdot\left[\br_j(t) - \br_j(0)\right]}\rangle$. The long wavelength limit of $F_s(\bq;t)$ is related to the mean-squared displacement, $F_s(\bq\to 0,t) = 1 - \frac{q^2}{6}\Delta r^2(t)$, and therefore the long time dynamics of $F_s$ at large wavelength allows us to obtain the diffusion constant. 

The standard mode-coupling approximation applies to equilibrium dynamics, though recent inroads~\cite{fuchs2009mode,szamel2016theory,liluashvili2017mode,szamel2019mode} pave the way for nonequilibrium extensions. The main technical difficulty in our case is the presence of transverse currents, which come in addition to the usual longitudinal ones. Within our own mode-coupling approximation for transverse forces, we obtain the memory kernels $(M_\parallel(\bq,t), M_\perp(\bq,t))$ encoding respectively longitudinal and transverse current-current correlations. The evolution of $F_s(\bq,t)$ is given by~\cite{federico-mct}
\begin{equation}\label{eq:mct}\begin{split}
\p_t F_s+T\mu q^2 F_s+ \mu\beta(1+\gamma^2)M_\perp*F_s = \\ -\left[\mu\beta (M_\parallel + M_\perp)+ \mu^2\beta^2(1+\gamma^2)[M_\perp* M_\parallel]
\right]*\p_t F_s ,
\end{split}
\end{equation}
where $*$ denotes a time-convolution. The functional expression of $M_\parallel$ is the same as in equilibrium~\cite{bengtzelius1984dynamics}, while
\begin{equation}
    M_\perp=T^2\rho_0 \int \frac{d\bk}{(2\pi)^3}\left( \frac{\bA\bq\cdot\bk}{\lvert \bA\bq\rvert}\right)^2c(k)^2 F_s(\bq-\bk,t)S(\bk,t) , \nonumber
\end{equation}
with $\rho_0$ is the number density, $S(\bk,t)$ the collective intermediate scattering function and $\rho_0 c(k) \equiv 1 - 1/S(k)$. The same matrix $\mathbf A$ as in our numerics is used. To close Eq.~\eqref{eq:mct} we need an equation of motion for $S(\bq;t)$. This equation, discussed in detail in \cite{federico-mct}, also predicts that the location of the mode-coupling transition is not influenced by the transverse currents, thus confirming the infinite-dimensional results. 

The zero-frequency mode of the memory kernel $\widehat{M}_{\alpha,i}=\lim_{q\to 0}\int_0^{+\infty}M_\alpha(q\be_i,t)\dd t$ controls the behavior of the diffusion constants, 
\begin{eqnarray}
    D_{\parallel, x}& = & T\mu \frac{1 + (1+\gamma^2)\beta \widehat{M}_{\perp,x}}{(1 + \beta \widehat{M}_{\parallel,x})(1 + \beta\widehat{M}_{\perp,x}) + \gamma^2\beta^2\widehat{M}_{\parallel,x}\widehat{M}_{\perp,x}} , \nonumber \\
D_{\parallel, z} &= &\frac{T\mu}{1+\beta\widehat{M}_{\parallel,z}},
\label{eq:diff}
\end{eqnarray}
with $D_{\parallel,y}=D_{\parallel,x}$. We note two consequences of working in finite dimension: $M_\perp\neq M_\parallel$ and $D_{\parallel,z}\neq D_{\parallel,x}$ (note that replacing both $M_\perp$ and $M_\parallel$ with $M$ in Eq.~\eqref{eq:diff} for $D_{\parallel,x}$ leads back to the infinite-dimensional expression Eq.~\eqref{eq:Ddinfinie}). Assuming that the system falls into a nonergodic regime below some transition temperature $T_\text{\tiny MCT}$, the memory kernels $M_\perp(t)$ and $M_\parallel(t)$ also saturate at a nonzero value at long times, and the longitudinal diffusion constants vanish. Whereas the location of the ergodicity breaking transition is independent of $\gamma$, the dynamics in the ergodic phase is not. In particular, assuming that $\widehat{M}_{\parallel,i}$ does not exceed its equilibrium counterpart, one can show that the longitudinal diffusion constants for $\gamma \neq 0$ are always larger than their equilibrium counterpart. If the diffusion constant is larger, the long time relaxation of $F_s$ is faster, and thus the value of the zero-frequency limit of the kernels is reduced, self-consistently demonstrating acceleration of the dynamics for $\gamma >0$. Remarkably, Eq.~\eqref{eq:diff} shows that the diffusion (quantified by  $D_{\parallel, z}$) along the $z$-direction is also indirectly accelerated by the coupling with the other directions. Overall, the mode-coupling calculation highlights interesting differences with the large $d$ limit, but the main results are in agreement. 

In conclusion, we found that the acceleration provided by transverse forces in a dense interacting system strongly depends on temperature, which comes as a surprise. The acceleration departs from a simple rescaling of the time, due to both interactions and emerging glassiness, which also lead to non-trivial asymptotic scaling with $\gamma$. Transverse forces begin to operate when the relaxation time of the system exceeds $\tau_\Sigma \sim {\tau_0}/{\gamma^2}$, but their efficiency decreases in deeply supercooled states leading instead to circular trajectories but only modest acceleration. This picture is corroborated by the behavior of the odd diffusivity, which is small as long as $\tau_\Sigma$ exceeds the relaxation rate of the system, but saturates to a finite value as the glass phase is approached. Our study resorts to a very local, and somewhat uninformed, way of driving the system out of equilibrium. In the more elaborate methods implemented in \cite{kapfer2017irreversible,michel2014generalized}, spatially extended and correlated moves are performed. It is a stimulating open question to find out how, when pushed towards glassiness, these methods compare with the minimal ones investigated here.

\acknowledgements  LB, FG and FvW acknowledge the financial support of the ANR THEMA AAPG2020 grant, along with several discussions with M. Michel, A. Guillin and W. Krauth. 
GS acknowledges the support of NSF Grant No.~CHE 2154241.

\bibliography{final_biblio}

\end{document}

\title{Supplementary Material to Sampling efficiency of transverse forces in dense liquids}

\author{Federico Ghimenti}

\affiliation{Laboratoire Mati\`ere et Syst\`emes Complexes (MSC), Université Paris Cité  \& CNRS (UMR 7057), 75013 Paris, France}

\author{Ludovic Berthier}

\affiliation{Laboratoire Charles Coulomb (L2C), Université de Montpellier \& CNRS (UMR 5221), 34095 Montpellier, France}

\affiliation{Yusuf Hamied Department of Chemistry, University of Cambridge, Lensfield Road, Cambridge CB2 1EW, United Kingdom}

\author{Grzegorz Szamel}

\affiliation{Department of Chemistry, Colorado State University, Fort Collins, Colorado 80523, United States of America}

\author{Fr\'ed\'eric van Wijland}

\affiliation{Laboratoire Mati\`ere et Syst\`emes Complexes (MSC), Université Paris Cité  \& CNRS (UMR 7057), 75013 Paris, France}

\date{\today}

\maketitle
\section{Escape rate}

In this Section we define the rate at which the system escapes its local configuration and we derive its expression for the transverse force dynamics. In the main text, the escape rate appears after Eq.~(1) when we discuss the various relevant time scales for our system.\\

Consider the motion of a particle in $d$ dimensions under the influence of transverse forces for an external potential $V$:
\begin{equation}
\frac{\dd\br}{\dd t}=-\mu(1+\gamma {\mathbf A})\p_\br V+\sqrt{2\mu T}\beeta 
\end{equation}
with $\left\langle \beeta(t) \otimes \beeta(t') \right\rangle = \mathbf{1} \delta(t-t')$. Since the noise is Gaussian, the probability $P(\br_i + \dot\br_i\dd t | \br_i, 0)$ of observing a transition from an initial position $\br_i$ to the position $\br_i + \dot\br_i \dd t$ in the infinitesimal time interval $\dd t$ is
\begin{equation}\label{eq:P}
    P(\br_i + \dot\br_i\dd t | \br_i, 0) \propto \exp\left\{-\frac{\dd t}{\mu} \left[\frac{1}{4 T}\left( \dot\br + \left(\mathbf{1} + \gamma\bA \right)\p_\br V\right)^2 - \frac{1}{2}\p_\br^2 V\right]\right\}
\end{equation}
If we set $\dot\br=0$ in Eq.~\eqref{eq:P} we obtain the probability that the particle remains in its initial configuration during the infinitesimal interval $\dd t$, i.e. \begin{equation}
    P(\br_i, \dd t | \br_i, 0) \propto \exp\left\{-\frac{\dd t}{\mu} \left[\frac{1}{4 T}\left(\left(\mathbf{1} + \gamma\bA \right)\p_\br V\right)^2 - \frac{1}{2}\p_\br^2 V\right]\right\}.
\end{equation}
This allows us to interpret the quantity with $\tau_0^{-1} \equiv \frac{1}{T \mu} \left\lvert \frac{1}{4 }\left( \left(\mathbf{1} + \gamma\bA \right)\p_\br V\right)^2 - \frac{T}{2}\p_\br^2 V\right\rvert$ as the rate at which the system escapes from its current configuration. The two terms in $\tau_0^{-1}$ have the following physical meaning: the first one expresses that nonzero forces make the system move away from a given configuration. The second term tells about the influence of the concavity of the potential on the escape process. 

We can average the escape rate over the steady-state Boltzmann distribution $\rho_\text{B} \propto e^{-\beta V}$. Using the fact that $\bA = -\bA^T$ and an integration by parts we get
\begin{equation}
    \left\langle\tau_0^{-1}\right\rangle_{\text{B}} = \frac{1}{4\mu T}\left[ \left\langle \left(\p_\br V\right)^2\right\rangle_\text{B} + \gamma^2 \left\langle \left(\bA \p_\br V\right)^2 \right\rangle_\text{B}\right].
\end{equation}
For $\gamma=0$ we recover the equilibrium result $\tau_{0,\text{eq}}^{-1} = \frac{1}{4T\mu}\left\langle \left(\p_\br V\right)^2\right\rangle_\text{B}$. If the potential is spherically symmetric, $\left\langle\left(\bA \p_\br V\right)^2 \right\rangle_\text{B} = \frac{1}{d}\lvert \lvert\bA\rvert \rvert^2_\text{F} \left\langle \left(\p_\br V\right)^2\right\rangle_\text{B}$ with $\lvert\lvert \bA \rvert\rvert_\text{F}^2 \equiv \sum_{i,j} A_{ij}^2$ the Frobenius norm of $\bA$. We therefore conclude that
\begin{equation}
    \tau_{0}^{-1} = \left(1 + \frac{1}{d}\gamma^2\lvert\lvert\bA \rvert\rvert^2_\text{F} \right) \tau_{0,\text{eq}}^{-1}.
\end{equation}
This result shows that transverse forces increase, through a rescaling, the escape rate of the system with respect to the equilibrium case. Note that in our infinite-dimensional calculation we have imposed that $||A||_\text{F}^2=d$ so that $\tau_0$ has a well-defined $d\gg 1$ limit.

\section{Connection between transverse forces and a lifted model}
In this Section we define a lifted model inspired by the active Ornstein-Uhlenbeck process and highlight its connection with transverse forces. 

We first recall the general definition of a \textit{lifted} Markov Chain. A system described by a degree of freedom $\br$ is said to be  lifted when it is endowed with an additional set of degrees of freedom $\bv$ and an irreversible (detailed balance-violating) Markovian dynamics in the newly defined extended space. The dynamics is constructed so that the steady state distribution for $\br$, after averaging over the variable $\bv$, is a desired target distribution. In our case the target distribution is the Boltzmann distribution $\rho_\text{B}\propto e^{-\beta V}$. If $\rho_{ss}(\br,\bv)$ is the steady state distribution of the lifted dynamics, we then need to impose $\int \dd \bv \rho_{ss}(\br,\bv) = \rho_\text{B}(\br)$. Here we consider as an example the following specific  lifted dynamics
\begin{equation}\label{eq:lAOUP}
    \begin{split}
        \dot\br &= -\mu \partial_\br V + v_0\bu +\sqrt{2T\mu}\beeta \\
        \dot\bu &= -\bu - \beta v_0 \p_\br V + \sqrt{2}\bchi.
    \end{split}
\end{equation}
Here $\bv = v_0 \bu$ plays the role of the lifting variable. One can check that $\rho_{ss}(\br,\bu) \propto \rho_\text{B}(\br)e^{-\frac{\bu^2}{2}}$ and that the dynamics is irreversible, since the entropy production rate reads 
\begin{equation}
\tau_\Sigma^{-1} =\frac{dv_0^2}{\mu T}+(\beta v_0)^2\langle(\p_\br V)^2\rangle_{\text{B}}>0.
\end{equation}
Interestingly, this is the same expression, up to an additive constant, as the one obtained after Eq.(1) in the main text.  We refer to Eq.~\eqref{eq:lAOUP} as a lifted AOUP ($\ell$AOUP). 

To establish an explicit connection with transverse forces, we consider the case of a $\ell$AOUP with temperature-dependent mobility, $\mu=T^{-1}$. If we introduce the extended variable $\bx = \left( \br, \bv\right)^T$ and the extended potential $U(\bx) = \beta V(\br) + \frac{\bu^2}{2}$ we obtain from Eq.\eqref{eq:lAOUP}
\begin{equation}
    \dot\bx = -(\mathbf{1}_{2d} + \gamma\bA)\p_\bx U + \sqrt{2}\bchi(t)
\end{equation}
with $\bA = \begin{bmatrix} \mathbf{0}_d & -\mathbf{1}_d \\ \mathbf{1}_d & \mathbf{0}_d\end{bmatrix}$ a $2d\times2d$ skew-symmetric matrix, $\bchi$ a Gaussian white noise with correlations $\left\langle \bchi(t) \otimes \bchi(t')\right\rangle = \delta(t-t')\mathbf{1}_{2d}$ and $\gamma\equiv v_0$. Upon considering the extended space of $\bx$ and rescaling the mobility of the system, we have been able to bring forth the presence of transverse forces in a lifted dynamics. 

\section{Acceleration at large $\gamma$}
In this Section we provide supporting evidence for the large $\gamma$ behavior of the diffusion constant.
\begin{figure}[h]
    \includegraphics[width=0.6\textwidth]{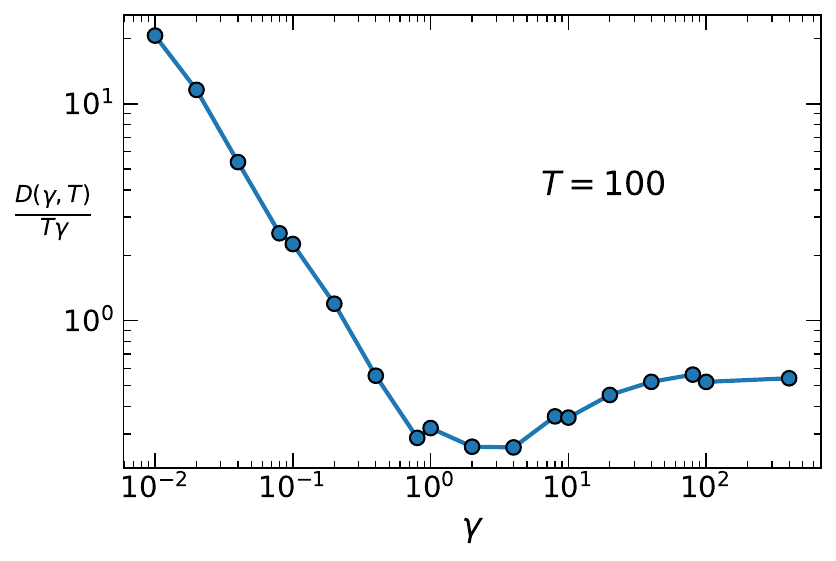}
    \caption{Behavior of the ratio $D(\gamma,T)/T\gamma$ measured for $T=100$ as a function of $\gamma$. For large values of $\gamma$, we see the scaling $D(\gamma,T)\sim \gamma$.}
    \label{fig:D_large_gamma}
\end{figure}
Fig.~\ref{fig:D_large_gamma} shows the behavior of the ratio $\frac{D(\gamma,T)}{\gamma T}$ as a function of $\gamma$, evaluated at $T=100$. For large values of $\gamma$ this quantity reaches a plateau, signaling the scaling $D(\gamma,T) \sim \gamma$. This result is also predicted from both the dynamical mean field theory and the mode coupling theory, as stated in the main text.
\section{Transverse forces in a harmonic well}

The goal of this Section is to investigate the properties of transverse forces for a particle in a two-dimensional harmonic well. We will study both the relaxation rate of the system and its odd diffusion. The results gathered here help to build up an intuition on the performance of transverse forces in systems characterized by caging effects such the ones investigated in the main text.

We consider the dynamics of a particle in a two-dimensional isotropic harmonic well of stiffness $k$ under the action of transverse forces:
\begin{equation}\label{eq:harmonic}
 \dot\br = -\mu k(\mathbf{1} + \gamma \bA)\br + \sqrt{2T\mu}\beeta(t)
\end{equation}
where $\beeta(t)$ is a Gaussian white noise with correlations $\left\langle \beeta(t) \otimes \beeta(t')\right\rangle = \mathbf{1} \delta(t-t')$. 

\subsection{Relaxation time}

To determine the relaxation time, we will map the Fokker Planck operator associated to the dynamics with transverse forces to a quantum mechanical problem. The Fokker-Planck operator $\Omega_\gamma$ reads
\begin{equation}
\Omega_\gamma = \mu \partial_\br \cdot \left[ (\mathbf{1} + \gamma \bA)\br + T \partial_\br\right].
\end{equation}
It governs the evolution of the probability distribution $\rho(\br,t)$, $\p_t \rho(\br,t)=\Omega_\gamma \rho(\br,t)$. The steady-state solution is the Boltzmann distribution $\rho_{\text{B}}\propto e^{-k\beta \frac{\br^2}{2}}$. The mapping to a quantum mechanical problem is performed by considering the operator
\begin{equation}
    H_\gamma \equiv -\rho_{\text{B}}^{-1/2}\Omega\rho_\text{B}^{1/2} = H_0 - \mu \gamma k \left(\bA\br\right) \cdot \p_\br
\end{equation}
where $H_0 = T\mu \left( -\bnabla + \beta k\frac{\br}{2}\right)\left(\bnabla + \beta k\frac{\br}{2}\right)$ is the Hermitian operator usually found for equilibrium dynamics and the second term is a skew-Hermitian operator. Its appearance is a consequence of injecting irreversible currents in the system through the transverse forces. 

To diagonalize $H_\gamma$, we first introduce a set of creation and annihilation operators 
\begin{equation}
    \begin{split}
        \ba &\equiv \sqrt{\frac{T}{k}}\left( \p_\br + \beta k \frac{\br}{2}\right) \\
        \ba^\dagger &\equiv \sqrt{\frac{T}{k}}\left( -\p_\br + \beta k \frac{\br}{2}\right)
    \end{split}
\end{equation}
that satisfy te commutation relations $\left[ a_i,a^\dagger_j\right]=\delta_{ij}$. This makes possible to rewrite the Hamiltonian $H_\gamma$ as
\begin{equation}
    H_\gamma = \mu k\ba^\dagger \cdot \begin{bmatrix} 1 & +\frac{\gamma}{2} \\ -\frac{\gamma}{2} & 1\end{bmatrix} \cdot \ba. 
\end{equation}
This operator can now be diagonalised using the change of basis
\begin{equation}
    \begin{split}
        \bbb &\equiv \frac{1}{\sqrt{2}}\begin{bmatrix} 1 & i \\ 1 & -i \end{bmatrix} \ba \\
        \bbb^\dagger &\equiv \frac{1}{\sqrt{2}}\begin{bmatrix} 1 & i \\ 1 & -i \end{bmatrix} \ba^\dagger 
    \end{split}
\end{equation}
which yields
\begin{equation}
    H_\gamma = k\mu \bbb^\dagger \begin{bmatrix} \left(1-i \gamma\right) & 0 \\ 0 & \left(1+i \gamma\right) \end{bmatrix} \bbb.
\end{equation}
It follows from the theory of the quantum harmonic oscillator that the spectrum of $H_\gamma$ is discrete and characterized by a two-dimensional vector $\bn = (n_1, n_2)^T \in \mathbb{N}^2$:
\begin{equation}
    \lambda_{\bn} = k\mu \left[n_1 + n_2 + i \gamma\left(n_2-n_1\right)\right]
\end{equation}
We see that the effect of transverse forces on the harmonic oscillator is a purely imaginary contribution to its spectrum, which does not change the relaxation time, but produces oscillations in the motion of the particle. In the next Section we will capture this swirling motion by computing the odd diffusion constant of the system.
\subsection{Odd diffusion}
In this Section we compute the odd diffusion in the harmonic well. The odd diffusion constant $D_\perp$ is 
\begin{equation}\label{eq:spectrum}
    D_\perp \equiv -\frac{\mu}{2} \int_0^{+\infty} \left\langle \dot\br(t) \cdot \bA \dot \br(0) \right\rangle
\end{equation}
where $\left\langle\ldots\right\rangle$ is the average of the initial position over the Boltzmann distribution and over the realisation of the noise $\beeta$. Note that we chose the sign of $D_\perp$ so that $D_\perp<0$ if the system undergoes a counterclockwise swirling motion. 

Using Eq.~\eqref{eq:harmonic} and the fact the realization of the noise $\beeta(t)$ are independent from $\beeta(0)$ and $\br(0)$ we get
\begin{equation}
    D_\perp = -\frac{\mu}{2}\int_0^{+\infty} k^2\left\langle \left(\mathbf{1} + \gamma\bA\right)\br(t) \cdot \bA \left(\mathbf{1} + \gamma\bA\right) \br(0)\right\rangle - \sqrt{2T} k \left\langle \left(\mathbf{1} + \gamma\bA \right)\br(t)\cdot\bA\beeta(0)\right\rangle.
\end{equation}
To proceed, we use the solution $\br(t)$ of Eq.\eqref{eq:harmonic}:
\begin{equation}
    \br(t) = \br(0) e^{-\mu k\left\langle \mathbf{1} + \gamma\bA\right)t} + \sqrt{2T\mu} \int_0^t \dd \tau e^{-\mu k\left(\mathbf{1} + \gamma \bA\right)(t-\tau)}\beeta(\tau)
\end{equation}
and the fact that $\left\langle \br(0) \otimes \br(0) \right\rangle_\text{B} = \mathbf{1}\frac{T}{k}$. Integration over time leads to
\begin{equation}
    D_\perp = -\frac{T\mu}{1+\gamma^2}\Tr \left[\left(\mathbf{1} - \gamma\bA\right)\left(\mathbf{1} - \gamma\bA\right)\bA\left(\mathbf{1} + \gamma\bA\right)\right] = -\gamma T\mu. 
\end{equation}
This is the expression for the odd diffusivity of the harmonic oscillator under transverse forces. We conclude with two remarks. First, the expression for $D_\perp$ is the same as the one found in an infinite dimensional liquid with transverse forces at the glass transition (Eq.~(10) of the main text with $\widehat{M}\to\infty$) or from mode coupling theory, providing a simple picture to explain the effect of transverse forces close to dynamical arrest. The second remark is that $D_\perp$ does not depend on $k$, the stiffness of the harmonic well. This is physically due to the following cancellation effect: the odd diffusivity scales as $F^2 \tau_\text{R}$, where $F^2$ is the average squared force acting on the particle and $\tau_\text{R}$ the relaxation time of the particle. Now, $F^2\sim k^2\left\langle \br^2 \right\rangle_\text{B} \sim k$, while $\tau_k \sim k^{-1}$, see Eq.\eqref{eq:spectrum}. The two factors cancel out, leaving $D_\perp$ independent from the well stiffness. 

\section{Dynamical Mean field Theory with transverse forces at low density}

In this Section we will address a low density expansion of dynamical mean field theory described in the main text in the case of a linear potential. This will lead to an explicit expression of the integrated memory kernel $\widehat M$, and thus to an explicit calculation of the efficiency $\frac{D(\gamma,T)}{D(0,T)}$ in the mean field theory. Here we will provide a sketch on how the expansion is implemented and describe the results it gives, while we refer the reader to \cite{federico-infinite} for a detailed derivation.

The basic idea is to expand the memory kernel of Eq. (7) of the main text in power of the density of the system $\rho\equiv \frac{N}{V}$:
\begin{equation}\label{eq:M_lowdensity}
    \mathbf{M} = \mathbf{M}^{(1)} + \mathbf{M}^{(2)} + \ldots
\end{equation}
The expression of $\mathbf{M}^{(n)} \sim O(\rho^n)$ can be self-consistently determined from the two-body process in Eq. (8), evaluated truncating the series in Eq.\eqref{eq:M_lowdensity} at the order $\rho^{n-1}$. In infinite dimension we can show that the $\mathbf{M}=\mathbf{1}M \approx \mathbf{1} M^{(1)}$ is independent from $\gamma$ \cite{federico-infinite}. This is a consequence of the infinite dimensional limit, where even for $\gamma \neq 0$ only the radial component of the two-particle process matters. The expressions for the longitudinal and the odd diffusion constants, $D(\gamma,T)$ and $D_\perp(\gamma,T)$ read
\begin{equation}
    \begin{split}
        D(\gamma,T) &\approx T\mu\frac{1+(1+\gamma^2)\beta\widehat{M^{(1)}}}{\left(1+\beta\widehat{M^{(1)}}\right)^2+\left(\beta\gamma\widehat{M^{(1)}}\right)^2} \\
        D_\perp(\gamma,T) &\approx -\gamma T\mu \frac{\beta \widehat{M^{(1)}}+(1+\gamma^2)\left(\beta\widehat{M^{(1)}}\right)^2}{\left(1+ \beta \widehat{M^{(1)}}\right)^2+\left(\gamma \beta \widehat{M^{(1)}}\right)^2}
    \end{split}
\end{equation}
In the case of a linear potential with interaction range $l$, $V(r)\equiv \epsilon \left\lvert \frac{r}{l}-1\right\rvert \theta\left(1-\frac{r}{l}\right)$ the memory kernel $\widehat{M^{(1)}}$ can be analytically determined. The calculation is the same as the one performed in \cite{manacorda2020numerical}. Its expression reads, upon introducing the rescaled packing fraction $\hat\phi \equiv \rho V_d \frac{l^d}{d}$, with $V_d$ the volume of a sphere of unit radius in $d$ dimensions
\begin{equation}\label{eq:M_lowphi_linpot}
    \beta\widehat{M^{(1)}} = \frac{\widehat\phi}{2}\frac{\beta^2\left(2+\beta\right)}{\left(1 + \beta \right)^3}.
\end{equation}
where we have used units such that $l=1$, $\epsilon=1$.

\begin{figure}
  \centering
  \includegraphics[width = 0.7\columnwidth]{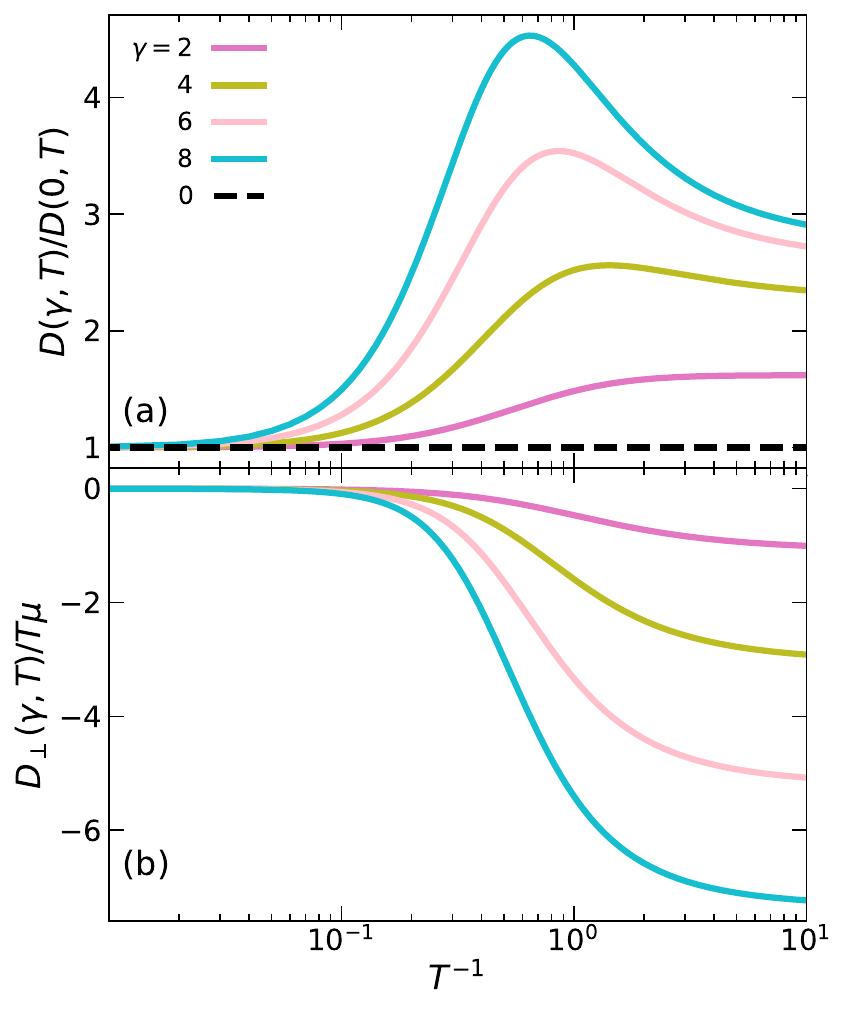}
  \caption{(a) Ratio between the diffusion constant $D(\beta, \gamma)$ in the presence of transverse forces and its equilibrium counterpart $D(\beta,0)$ for different values of the strength $\gamma$ of the nonequilibrium drive, as a function of the inverse temperature $\beta\equiv T^{-1}$. (b) Odd diffusivity in the presence of transverse forces. In  both panels, the memory kernel used is the one obtained via a low density expansion for the case of a linear potential. Its expression is given in Eq. \eqref{eq:M_lowphi_linpot}, with $\widehat\phi=1$.}
  \label{fig:Drel_lowphi}
\end{figure}

Figure \ref{fig:Drel_lowphi} shows the efficiency $\frac{D(\gamma,T)}{D(0,T)}$ and the odd diffusion constant $D_\perp(\gamma,T)$ calculated using the low density expansion mentioned above for several values of $\gamma$ as a function of $\beta$. We observe a very strong qualitative similarity with the high temperature behavior of the efficiency found in Fig. 3(a) of the main text. The efficiency evolves nonmonotonically with $\beta$ for the larger values of $\gamma$, while the $D_\perp$ grows steeply from $0$ to a nearly constant value. The growth of $D_\perp$ happens around the point of maximal efficiency, and absolute value of the plateau value of $D_\perp$ grows with $\gamma$. 
\bibliography{final_biblio}